\documentclass[preprint,12pt]{elsarticle}
\usepackage{bm}
\usepackage{amsmath}
\usepackage{amssymb}
\usepackage{xcolor,soul}
\usepackage{siunitx}
\usepackage{booktabs}
\usepackage{upgreek}
\usepackage{pdfpages}
\newcolumntype{P}[1]{>{\centering\arraybackslash}p{#1}}
\DeclareSIUnit\angstrom{\text {Å}}

\newcommand{\wei}[1]{#1}   % <-- submission version (no highlighting)

\journal{ArXiv}

\begin{document}

\begin{frontmatter}

%\title{Forward-Backward Asymmetry in the Mobility of Jogged Dislocations in Face-Centered Cubic Nickel \hl{better title?}}
%\title{Asymmetric Dislocation Mobility Induced by Jogs}

% compare with -- Grain boundaries are brownian ratches
% -- dislocations are brownian ratches too
% -- can we work in the word ratches
\title{Crystal Dislocations as Atomic Scale Ratchets}

\author[1]{Wu-Rong Jian\fnref{equal}}

\author[1]{Yifan Wang\fnref{equal}}%\corref{corr-author}}
%\ead{yifan.wang@oist.jp}

\author[1]{Wei Cai\corref{corr-author}}
\ead{caiwei@stanford.edu}

%\address[1]{State Key Laboratory of Subtropical Building and Urban Science, South China University of Technology, Guangzhou, Guangdong 510640, P. R. China}
%\address[2]{Department of Engineering Mechanics, South China University of Technology, Guangzhou, Guangdong 510640, P. R. China}
\address[1]{Department of Mechanical Engineering, Stanford University, Stanford CA, 94305, USA}
%\address[4]{Okinawa Institute of Science and Technology, Onna-son, Okinawa 904-0495, Japan}

\fntext[equal]{Both authors contributed equally to the paper}
\cortext[corr-author]{Corresponding author}

\date{\today}

\begin{abstract}

% Wei (2026/05/19)
The symmetry of a system's response to external stimuli is a fundamental
concept in physics and materials science.
At the microscopic scale, breaking this symmetry to achieve a rectified response is exceptionally difficult to engineer and remains rare in nature.
Conventional micromechanics models of crystalline solids often assume a symmetric response to applied stress, where reversing the load simply inverts the direction of defect velocity without altering its magnitude.
In this work, we report an atomic-scale, geometry-rooted mechanism that breaks this symmetry. Molecular dynamics simulations of face-centered cubic nickel reveal that dislocations containing atomic-scale jogs exhibit asymmetric mobility under opposite applied stresses: reversing the loading direction triggers significantly higher drag. \wei{This asymmetry arises from the coupling of two internal variables with different transformation parity: a non-affine displacement of an atom at the jog core, and a strain-like tensor associated with the advance of the dislocation.}
%
%Atomistic energy barrier calculations confirm that this coupling induces an asymmetric activation barrier response under opposing stresses.
%
Because jogs are ubiquitous structures in plastic deformation, this discovery challenges classical descriptions of plastic deformation mechanisms, with direct implications for cyclic creep, and opens new pathways for defect engineering to enhance fatigue resistance.
\end{abstract}

\end{frontmatter}

%%%%
%%%%
\section{Introduction}
Characterizing how a system responds to external stimuli is a foundational concept across physics and materials science~\cite{onsager_reciprocal_1931, green_markoff_1954, kubo_statistical-mechanical_1957, kondepudi_modern_2014}.
%
% \hl{(linear response theory of non-equilibrium thermodynamics - Prigorgin, Onsager, Green-Kubo)}
Typically, these response functions are symmetric: reversing the direction of a stimulus while maintaining its magnitude results merely in an equal and opposite response.
%
%The symmetry of a system's response to external stimuli is a fundamental concept in physics and materials science. 
%
%Many systems can be characterized by a symmetric response: reversing the direction of the stimulus while keeping its magnitude constant, results only in a sign change of the response without altering its magnitude.
%
In the macroscopic world, asymmetric or rectified behaviors are readily engineered through geometric constraints, such as structural ratchets and one-way valves, which convert symmetric, alternating inputs into unidirectional motion.
%In the macroscopic world, asymmetric responses can be achieved through engineered mechanisms.
%
%For instance, rachets can convert symmetric alternating inputs into directional motion,
%
%similar designs include diode and single-direction valve.
%
However, as the spatial scale approaches the atomic levels, engineering such asymmetrical behavior becomes exceptionally challenging, and its spontaneous occurrence remains rare in nature.
%
%However, as the scale is reduced, achieving this asymmetric (or racheting) responses become increasingly challenging in engineering and rare in nature.
%
Consequently, uncovering intrinsic microscopic mechanisms that inherently break this symmetry to produce a rectified output is of profound interest to both scientists and engineers~\cite{smoluchowski_experimentell_1912, feynman_feynman_1963, bier_brownian_1997, fornes_brownian_2021}.
%
%Consequently, the discovery of microscopic mechanisms that inherently break symmetry to produce rectified output remains a core fascination for scientists and engineers~\cite{smoluchowski_experimentell_1912, feynman_feynman_1963, bier_brownian_1997, fornes_brownian_2021}.% \hl{[Ref - ratchet mechanisms, Feynmann]}

A notable example of microscopic asymmetry is found in biology, where motor proteins such as kinesin convert non-directional stimuli into directed motion to power vital functions~\cite{schliwa2003nature}.
%
% and actin cytoskeleton convert.
%
%(Take a look at the PNAS 10.1073/pnas.1414184111 paper)
%
%many proteins (such as kinesin) act as molecular-scale machines that convert non-directional stimuli into directed motion, supporting vital biological functions~\cite{schliwa2003nature}.
%
This inspires a compelling question in materials science: Can analogous ``molecular machines'' occur naturally within crystalline solids, and what is the minimum scale required for such behavior?
Historically, asymmetric atomic motion in crystals has rarely been observed.
%The observation of asymmetric atomic motion in crystals has historically been rare.
%
The highly symmetric, periodic arrangement of atoms in a perfect lattice strongly favors a symmetric response.
%
%This scarcity is attributed to the fact that the highly symmetric, periodic arrangement of atoms in perfect crystals strongly favors a symmetric response.
%
While defects break this symmetry, they are typically considered too small to host a functional ratcheting mechanism.
%
%Although crystalline defects break these symmetries, they are often considered to be too small to accommodate the ratcheting mechanism needed for an asymmetric response.
%
%
Recently, however, certain grain boundaries, two-dimensional planar defects, have been shown to exhibit asymmetric mobility~\cite{qiu2024science}. 
%
%Recently, it has been shown that certain grain boundaries (two-dimensional, i.e. planar, defects in crystals) can exhibit asymmetric mobility \cite{qiu2024science}.
%
This discovery naturally motivates the inquiry into whether directional rectification can be achieved in even smaller, lower-dimensional defects within the crystal lattice.
%
%This finding motivates the inquiry into whether asymmetric mobility can be found in even smaller defects within crystals.

Moving down in dimension, dislocations are one-dimensional (line) defects within the crystal grains, representing structural features smaller than grain boundaries.
%, and thus representing defects smaller than the grain boundaries.
%
The motion of these line defects under applied stress is the primary mechanism governing the strength and plastic deformation of most crystalline materials~\cite{hull_introduction_2011, anderson_theory_2017, cai_imperfections_2016}.
% \hl{[Add Ref - books - Hirth Lothe, Nix Cai]}
%
%The motion of dislocations in response to the applied stress is the primary mechanism controlling the strength and plastic deformation behavior of most crystalline materials~\cite{bacon2009dislocation}. %\hl{[Ref - books - Hull Bacon, Hirth Lothe, Nix Cai]}
%
The simplest case of dislocation mobility is glide on a
slip plane in face-centered cubic (FCC) metals, such as copper, nickel, and aluminum.
At low to moderate stress levels, where the dislocation velocity remains well below the speed of sound, this motion is characterized by linear mobility governed by viscous phonon drag~\cite{blaschke2019jpcs}.
%
%This motion is often characterized by a linear mobility at stresses low enough that the dislocation velocity is far below the speed of sound.
%
%This mobility arises from viscous drag exerted by lattice phonons~\cite{blaschke2019jpcs}.
%
An asymmetric glide response in FCC metals has never been anticipated. Consequently, historical reports of dislocation mobility (from both atomistic simulations~\cite{cho2017ijp} and experimental measurements~\cite{greenman1967jap,jassby1970pm}) uniformly omit any consideration of the direction of motion, implicitly assuming the response is invariant to the sign of the applied stress.
%
%The notion of asymmetric 
%%\textcolor{red}{WJ: here, should be ``symmetric''?}
%glide motion of dislocations in FCC metals was never suspected; 
%%(\hl{possibility of being asymmetric was never thought of} -- ask AI to fix this?);
%to the extent that published reports of dislocation mobility (whether derived from atomistic simulations~\cite{cho2017ijp} or experimental measurements~\cite{greenman1967jap,jassby1970pm}) have historically omitted any mention of the direction of motion, implicitly assuming the response is independent of the sign of the applied stress.
%
% DRAFT (2026/7/26) -- BCC disclaimer, item 1 of the 2026/7/9 to-do list.
% Kept to a single sentence: acknowledge the BCC case without side-tracking
% the main story.  See plan/plan_edit_round1.md.
% NOTE: [REFS NEEDED] -- no BCC screw / non-Schmid entry in ref.bib yet.
%       Candidates (UNVERIFIED): Duesbery & Vitek (Acta Mater 1998);
%       Ito & Vitek (Phil Mag A 2001); Groger, Bailey & Vitek (Acta Mater
%       2008); Christian (Metall Trans A 1983).  Wei/Yifan to confirm.
Asymmetric \wei{motion (through kink-pair nucleation)} is known for screw dislocations in body-centered cubic (BCC) metals 
\wei{due to twinning--anti-twinning asymmetry of the BCC crystal}~\cite{maresca_screw_2018}.
%\yf{due to asymmetric kink-pair nucleation barriers}~\cite{maresca_screw_2018}.
%
\wei{However, because the kink-pair nucleation is not confined to a plane}~\cite{bulatov_kink_1997}, \wei{there is no precise ``forward'' or ``backward'' direction for the screw dislocation motion, and the asymmetry does not give rise to a clear ratchet behavior during cyclic loading.}
%\yf{However, kink pairs break the screw dislocation into small segments. None of their motion is confined to a single glide plane with reversed stress.
%
%As a result, the segmented screw dislocation nucleate and move randomly with cyclic loading}~\cite{bulatov_kink_1997}, \yf{hardly be seen as a whole mechanism that can carry ratchet behavior.}
%
% Reversing the stress changes the glide plane, and such segmented dislocations can hardly be treated as a whole ratchet  as well as the direction and no ratchet results.

However, during plastic deformation, moving dislocations on intersecting planes frequently collide, generating atomic-scale steps called jogs along the dislocation line.
%However, during plastic deformation, moving dislocations on different planes frequently intersect.
%  
%Such intersections often result in the formation of jogs, which are atomic-sized steps, on the dislocation line.
%
Through molecular dynamics (MD) simulations in FCC nickel, we found that dislocations containing these jogs can exhibit a distinct asymmetric mobility, where their speed varies significantly with the direction of motion.
%
%meaning that their speed depends on the direction of motion.
%
As a consequence, applying a symmetric cyclic load (such as high-cycle fatigue or ultrasound) results in a net directional drift of the dislocation line, a clear manifestation of mechanical rectification.
%
%As a consequence, we show that under symmetric cyclic loading (such as high-cycle fatigue or ultrasound), the jogged dislocation moves persistently in one direction (rectification).
%
%
Our results suggest that this asymmetric mobility is a general characteristic of jogged dislocations, whereas symmetric mobility occurs only for specific line orientations.
Because jogs are a natural and common consequence of plastic deformation, these asymmetric dynamics represent the norm, rather than the exception, in crystalline solids undergoing deformation.
%
%Given that jogs are a natural and common consequence of plastic deformation, jogged dislocations and asymmetric mobilities are the norm, not the exception in most crystalline materials undergoing deformation.

The underlying mechanism driving the asymmetric mobility is rooted in the local atomic structure, where jog translation requires the thermally-activated jump of a specific atom into an adjacent vacant site.
While a positive stress monotonically reduces the activation energy barrier for this atomic jump, a negative stress induces a non-monotonic response, initially decreasing but subsequently increasing the energy barrier for the reverse jump.
\wei{We demonstrate that this counterintuitive asymmetry is mechanically permissible because jog migration involves two internal variables with different transformation parity: a non-affine displacement of the atom at the jog core, and a strain-like tensor associated with the advance of the dislocation, leading to an unusual symmetry-breaking situation.}
%
%We show that such an asymmetry, although counter-intuitive, is allowed when the applied stress (a second-order tensor) is coupled with an atomic displacement (a vector), leading to an unusual symmetry-breaking situation.
%
Consequently, the local atomic structure at the dislocation jog functions as an intrinsic ``lattice ratchet'' that rectifies alternating applied stress.
%
%As a result, the atomic structure at the dislocation jog constitutes a ``lattice ratchet'' that responds asymmetrically to alternating stresses.
%
This finding is of fundamental importance for understanding the plasticity of crystalline solids, uncovering a new atomic-scale mechanism that directly influences macroscopic fatigue and cyclic deformation behavior~\cite{suresh_fatigue_1998}.
%
%This finding is fundamentally important for understanding the plastic deformation of crystalline materials (including the majority of metals and alloys) and reveals a new atomic-scale mechanism that directly influences fatigue behavior under cyclic stress~\cite{suresh_fatigue_1998}.%\hl{[Ref - fatigue - Suresh]}

\section{Results}

\begin{figure}[!ht]
    \centering
    \includegraphics[width=1.0\linewidth]{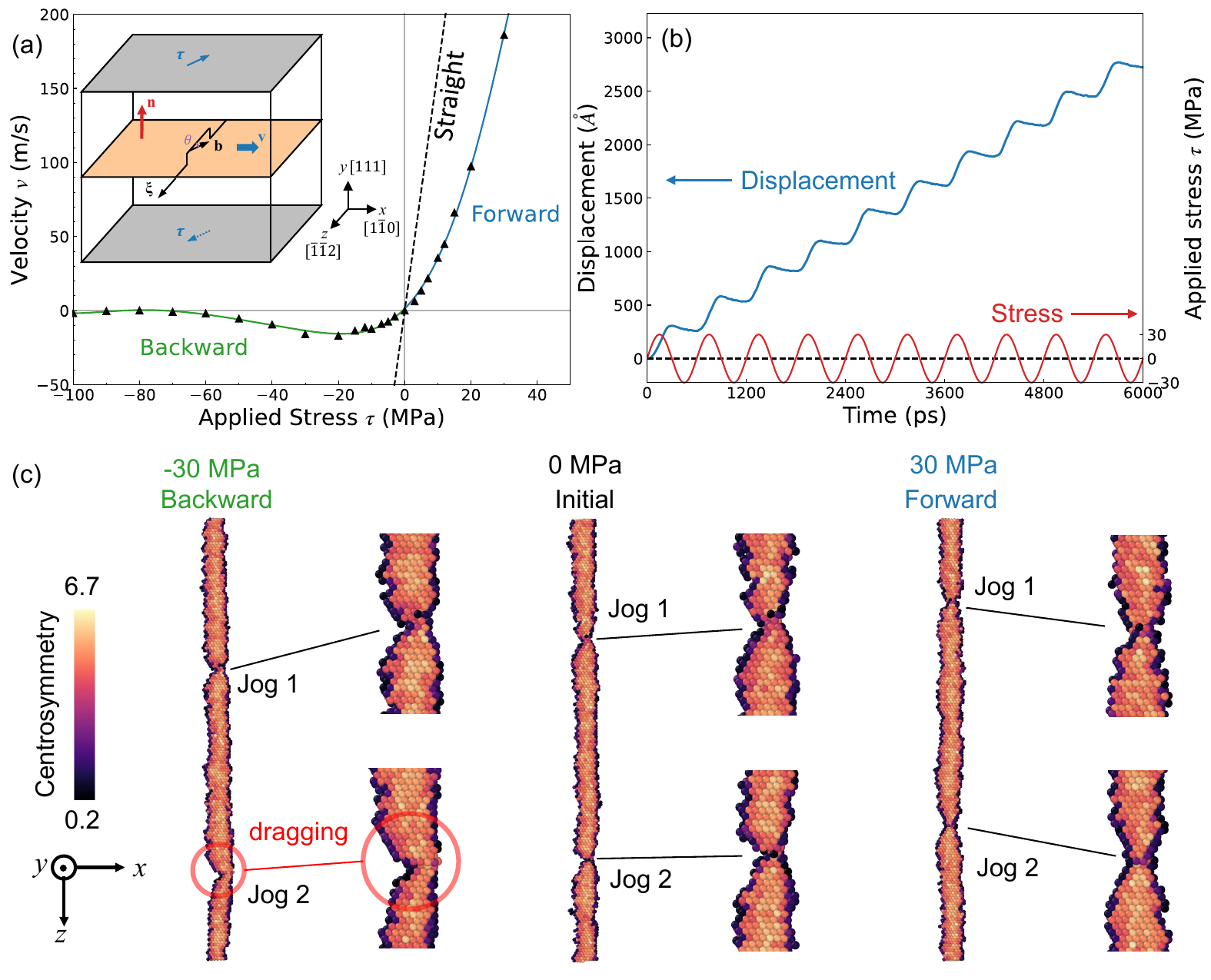}
    \caption{
    %\hl{(Shrink figure caption)}
    (a) Velocity of jogged (solid line) and straight (dashed line) dislocation with character angle $\theta = \qty{30}{\degree}$ as a function of applied stress $\tau$.
    The inset shows a schematics of the simulation cell containing a jogged dislocation with Burgers vector $\mathbf{b}$ and line direction $\boldsymbol{\xi}$.
    The dislocation moves on its glide plane with normal vector $\mathbf{n}$.
    The arrows on the top and bottom surfaces indicate the direction of applied stress $\tau$ when it is positive, which causes the dislocation to move in the positive $x$-direction (forward).
    (b) Displacement-time curve of the dislocation motion under sinusoidal cyclic loading at a stress amplitude of $\qty{30}{MPa}$.
    % \textcolor{red}{Double check the 30 MPa of stress magnitude.}
    %
    (c) Atomic structure of the jogged dislocation under backward loading ($\tau = -\qty{30}{MPa}$), no loading ($\tau = \qty{0}{MPa}$), and forward loading ($\tau = \qty{30}{MPa}$).
    The atoms are colored with the centrosymmetry parameter 
    %(12 neighbors, FCC) 
    using the OVITO software package~\cite{stukowski2009msmse}, showing the dislocation line containing two jogs (1 and 2).
    %
    % \hl{(Upload both 30 MPa and 20 MPa plots.)}
    % \hl{(Perhaps in (b) right y-axis remove all ticks at 60, 100, 200, 300, 400.)}
    }
    \label{fig:mobility}
\end{figure}
We created an atomistic structure of a $\theta=\qty{30}{\degree}$ mixed dislocation with two unit jogs (see Methods) in a simulation cell of $\qtyproduct{240x150x360}{\angstrom}$ in the $[1\bar{1}0]\times[111]\times[\bar{1}\bar{1}2]$ directions, as illustrated in the inset of Fig.~\ref{fig:mobility}(a).
Under the applied stress $\tau$ along the Burgers vector $\mathbf{b}$ direction on the slip plane, the dislocation line moves along the $x$-direction.
The positive $\tau$ direction (illustrated in the inset) drives the dislocation moving in the positive $x$ direction, while the negative $\tau$ direction drives the dislocation moving in the negative $x$ direction.
Fig.~\ref{fig:mobility}(a) plots the velocity $v$ of the jogged 30$^\circ$ dislocation as a function of the applied stress $\tau$ (solid line),
where the positive and negative $\tau$ represents forward and backward loading directions, respectively.
The presence of jogs slows down the dislocation compared to the straight dislocation without jogs (dashed line).
More significantly, the $v(\tau)$ curve for the jogged dislocation lacks the symmetry exhibited by the straight dislocation.
Specifically, the magnitude of the dislocation velocity in the backward direction is substantially lower than that in the forward direction at the corresponding stress magnitude, over a wide range of stress magnitudes (from $\qtyrange{10}{100}{MPa}$).
Mobility asymmetry of this type was never suspected before for dislocations in FCC metals.
%
%%%%%%%%%%%%%%%%% temporarily not include in the arxiv version
% In addition to $\qty{30}{\degree}$ case, we also observed asymmetrical mobility for jogged $\qty{60}{\degree}$ dislocations as well \hl{(see Supplementary Fig.~xx)}.
%%%%%%%%%%%%%%%%%%%%%%%%%%%%%%%%%%%%%%%%%%%%%%%%%%%%%%%%%%%%%%

Over the stress range considered here (from $\qtyrange{0}{100}{MPa}$), the $v(\tau)$ relation in the forward direction is non-linear,
indicating a thermally-activated mechanism at low stress,
similar to our previous findings on jogged edge dislocations~\cite{jian2026acta}.
Nonetheless, the $v(\tau)$ relation is monotonic, as expected.
In contrast, the velocity-stress relation in the backward direction is not only non-linear but also non-monotonic.
The dislocation velocity begins to decrease as the applied stress magnitude exceeds $\qty{20}{MPa}$, rendering the dislocation nearly immobile at $\qty{100}{MPa}$.
This surprisingly non-monotonic $v(\tau)$ behavior in the backward direction leads to a pronounced forward-backward asymmetry in the mobility of jogged mixed dislocations.

To assess the mechanical rectification of this asymmetric mobility behavior under cyclic loading conditions,
we perform a long MD simulation where the jogged dislocation was subjected to a sinusoidal shear stress with a $\qty{30}{MPa}$ amplitude and zero mean,
shown as the red curve in Fig.~\ref{fig:mobility}(b).
This stress level is relevant to conditions encountered in high-cycle fatigue or ultrasound.
The blue solid curve in Fig.~\ref{fig:mobility}(b) shows the displacement of the jogged 30$^\circ$ dislocation as a function of time.
During each half cycle where the applied stress is positive, the dislocation moves forward by approximately $\qty{307}{\angstrom}$.
In each half cycle with negative applied stress, the backward dislocation motion is significantly smaller, at about $\qty{-50}{\angstrom}$. %\textcolor{red}{WJ: use ``approximately'' instead of ``less than''? The value is $\qty{49.7}{\angstrom}$} $\qty{50}{\angstrom}$.
The observed dislocation displacement at each cycle agrees well with the expectation from the mobility values shown in Fig.~\ref{fig:mobility}(a) obtained from steady-state MD simulations (see {Supplementary Fig.~S8}). 
%
% (see \hl{Supplementary Fig.~S3}). % \hl{I don't find it in Supplementary Materials.}
%
Overall, the dislocation decisively moves in the forward direction with each stress cycle, even though the mean applied stress is zero.

Fig.~\ref{fig:mobility}(c) shows the shape of the dislocation at equilibrium under zero loading, as well as during steady-state motion under forward $(\tau = \qty{30}{MPa})$ and backward $(\tau = \qty{-30}{MPa})$ loading conditions.
Two jogs can be identified from these plots, with Jog 2 appearing somewhat more constricted than Jog 1, especially during the backward motion of the jogged dislocation.
Our analysis showed that Jog 1 is highly mobile and has negligible effect on dislocation motion, while Jog 2 exerts significant drag on dislocation motion.
The different behaviors of the two jogs are consistent with our previous findings on jogged edge dislocations~\cite{jian2026acta, wang2025scripta}.
The dissociated dislocation structures near the two jogs are shown in {Supplementary Fig.~S7}.
The drag effect of Jog 2 is especially pronounced under backward loading, and is solely responsible for the mobility asymmetry reported above.

\section{Discussion}

\subsection{Atomic mechanism of the asymmetric mobility}

\begin{figure}[!ht]
    \centering
    \includegraphics[width=0.7\linewidth]{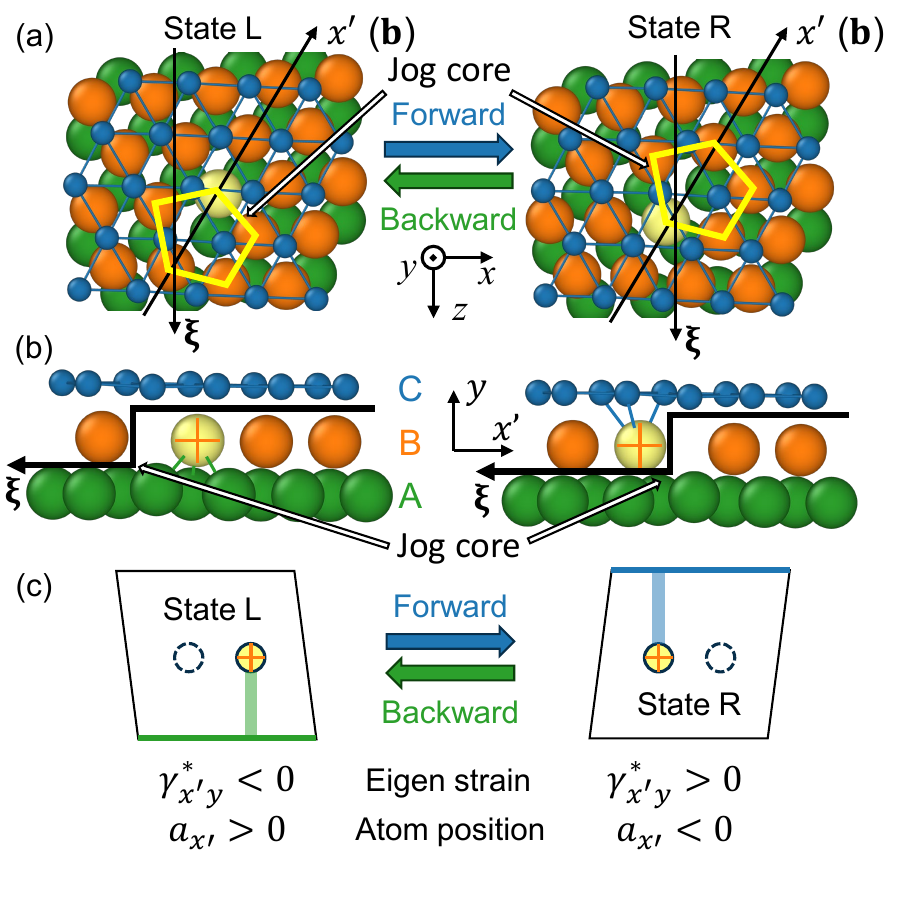}
    \caption{
    (a) Energy minimized atomic structures near Jog 2 of the two states (\emph{L} and \emph{R}) before and after jog motion, viewing from the top of the slip plane $x$-$z$.
    %
    %The green dislocation lines are obtained from DXA using OVITO~\cite{stukowski2012msmse}.
    % \textcolor{red}{Remove the green dislocation line. In supplementary, we can show dislocation lines with more atoms.}
    % changed to black solid line
    %
    $\pm x'$ is the jog motion direction,
    %
    %\textcolor{red}{this only corresponds to ``state L to state R'', forward motion of jogged dislocation? We cannot say this is the jog motion direction.}
    % This is the jog motion direction
    %
    aligned with the Burgers vector $\mathbf{b}$ direction.
    The jog core is marked as a yellow pentagon and the highlighted
    % \textcolor{red}{yellow? Add a small star inside.} 
    atom with the cross mark is the \emph{core atom} that moves in the opposite direction of the jog motion. % \textcolor{red}{jog motion?}.
    (b) Side view ($x'$-$y$ plane) of the atomic structure of Jog 2.
    The dislocation line is illustrated as a black arrowed line.
    The coupling between the \emph{core atom} and the neighboring layers is represented as three colored struts.
    (c) Simplified schematic of the two states, where
    \emph{State L} corresponds to the \emph{core atom} position $a_{x'} > 0$,
    eigenstrain $\gamma^*_{x'y} < 0$, and
    \emph{State R} corresponds to $a_{x'} < 0$, $\gamma^*_{x'y} > 0$.
    The solid circle indicates the location of the `core atom,' while the dashed circle is the location of the `core atom' in the other state.
    % \textcolor{red}{Do we need to explain what the solid and dashed circles are?}
    %\yf{(Refer to see Supplementary Fig.~S9.)}
    }
    \label{fig:mechanism}
\end{figure}

To understand the underlying mechanism responsible for this surprising forward-backward asymmetry in dislocation mobility,
we analyze the atomic structure at Jog 2 that induces asymmetrical drag.
%under forward and backward loading.
%with opposite loading.
%
We obtained the two neighboring atomic configurations of the jog moving by one Burgers vector in the MD simulation,
and relaxed them without applied stress to obtain the energy minimized configurations \emph{State L} and \emph{State R}. % \hl{(in large cell MD simulations, but later for NEB we used small cell)}.
Fig.~\ref{fig:mechanism}(a) shows the atomic structures near Jog 2 viewed down the slip plane (i.e. the $x$-$z$ plane),
%along the slip plane normal in -$y$ direction, 
and (b) shows the side view, i.e. on the $x'$-$y$ plane, where $x'$ is the Burgers vector direction.
We show only three stacking layers of atoms (\emph{A}, \emph{B}, \emph{C}) that are relevant to the jog structure.
Along the dislocation line direction $\boldsymbol{\xi}$, the unit jog brings the dislocation line (black lines in (b)) down by one slip plane,
i.e., from the slip plane between \emph{Layer B} and \emph{Layer C} to that between \emph{Layer A} and \emph{Layer B}. 
% \textcolor{red}{This sentence is confusing. Maybe try to rephrase it?}

The core structure of Jog 2 corresponds to an atomic vacancy in \emph{Layer B}, surrounded by a ring of 5 atoms marked as a yellow pentagon in Fig.~\ref{fig:mechanism}(a).
\wei{(For more details see Supplementary Fig.~S9.)}
The jog motion corresponds to the shift of the pentagon along the Burgers vector direction ($x'$ direction).
This is accomplished by a highlighted yellow atom, called the core atom below, jumping into the vacancy.
%
%while the corresponding atomic motion is the highlighted yellow \emph{core atom} moving in the opposite direction.
%
For example, in Fig.~\ref{fig:mechanism}(b), if the core atom in \emph{State L} jumps to the left, it results in the jog moving to the right, arriving at \emph{State R},
corresponding to the forward motion of the jogged dislocation.
%
% \textcolor{red}{and corresponding to forward motion of the jogged dislocation?}.
%
Conversely, if the core atom in \emph{State R} jumps to the right, the jog moves to the left, arriving at \emph{State L}
corresponding to the backward motion of the jogged dislocation.
%
% \textcolor{red}{and corresponding to backward motion of the jogged dislocation?}.
%
Fig.~\ref{fig:mechanism}(b) also shows that the dislocation slip plane at the marked core atom is different at the two states.
In \emph{State L}, the slip plane (which coincides with the black line in this view) is above the core atom, i.e. between \emph{Layer C} and \emph{Layer B} where a stacking fault exists.
Consequently, in \emph{State L} the core atom binds strongly with \emph{Layer A} below it, as indicated by the three green segments in Fig.~\ref{fig:mechanism}(b).
In \emph{State R}, however, the slip plane is below the core atom, i.e. between \emph{Layer B} and \emph{Layer A} where a stacking fault exists.
Consequently, in \emph{State R} the core atom binds strongly with \emph{Layer C} above it, as indicated by the three blue segments in Fig.~\ref{fig:mechanism}(b).
%
% \textcolor{red}{We need one sentence to clarify the forward motion involves the jog motion, the core atom motion and the dislocation motion.}

Fig.~\ref{fig:mechanism}(c) shows diagrams depicting the key differences between the two states.  First, the difference in the location of the core atom can be described by a vector $\boldsymbol{a}$ along the $x'$-direction.
Relative to a suitable reference, $a_{x'} > 0$ in \emph{State L} and $a_{x'} < 0$ in \emph{State R}.
Because the motion of the jog causes the entire dislocation to move and produce plastic strain, \emph{State L} and \emph{State R} also correspond to different eigenstrains $\boldsymbol{\gamma}^{*}$, which is the ``resting'' strain of the entire simulation cell under zero stress.
Because a positive stress $\tau_{x'y}$ promotes the transition from \emph{State L} to \emph{State R}, we can conclude that (relative to a suitable reference), $\gamma^{*}_{x'y} < 0$ in \emph{State L} and $\gamma^{*}_{x'y} > 0$ in \emph{State R}.

Because $a_{x'} \, \gamma^{*}_{x'y} < 0$ in both \emph{State L} and \emph{State R}, the jogged dislocation thus produces a coupling between a vector $\boldsymbol{a}$ and a second order tensor $\boldsymbol{\gamma}^{*}$. 
This coupling is quite peculiar in that, although the two resulting states shown in Fig.~\ref{fig:mechanism}(c) look symmetric, they are not.
For example, if we apply an $\qty{180}{\degree}$-rotation within the plane of the page to \emph{State L}, vector $\boldsymbol{a}$ will reverse sign ($a_{x'}$ will change from positive to negative).
However, the second-order tensor $\boldsymbol{\gamma}^*$ will remain unchanged ($\gamma^{*}_{x'y}$ stays negative); hence the result is not \emph{State R}.
\wei{Stated more precisely, jog migration involves two internal variables of different transformation parity: \mbox{$\boldsymbol{a}$}, a non-affine displacement of the core atom, is odd under this rotation, whereas \mbox{$\boldsymbol{\gamma}^{*}$}, a strain-like tensor associated with the advance of the dislocation, is even.
It is this mismatch in transformation parity of the two coupled variables that makes \emph{State L} and \emph{State R} physically distinct while appearing geometrically similar.}
These symmetry considerations provide some first hints at the origin of the mobility asymmetry of the jogged dislocation.

\begin{figure}[!ht]
    \centering
    \includegraphics[width=0.8\linewidth]{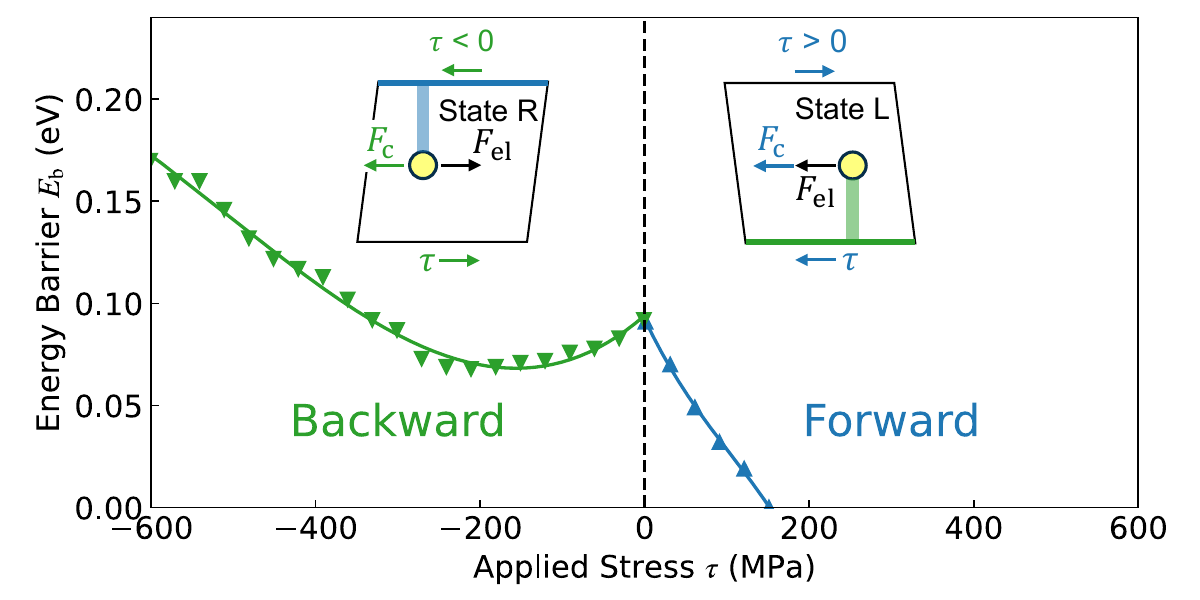}
    \caption{
        Energy barriers of forward and backward mechanisms 
        %\textcolor{red}{of jogged dislocation} 
        at different applied stresses.
        The corresponding inset schematics demonstrate the force analysis of the forward and backward mechanisms.
    }
    \label{fig:energy_barrier}
\end{figure}

To provide more concrete answers to the origin of the mobility asymmetry, we calculated the energy barriers between \emph{State L} and \emph{State R} as a function of stress in both forward and backward directions.
The energy barriers shown in Fig.~\ref{fig:energy_barrier} were calculated by the nudged elastic band (NEB)~\cite{henkelman2000jcp} method ({see Methods}).
In the forward direction ($\tau_{x'y} > 0$), the behavior appears to be consistent with a typical stress-driven thermally activated process~\cite{mcdowell2023jms, kuykendall2020jmps}, where the energy barrier decreases with increasing applied stress.
In the backward direction ($\tau_{x'y} < 0$), however, the energy barrier first decreases with increasing stress magnitude (but with a smaller slope), and then increases for stress magnitude above $\qty{200}{MPa}$.
%
%The energy barrier increases up to $\qty{0.18}{eV}$ under $\qty{600}{MPa}$ backward stress, which makes the backward motion almost impossible at the simulation temperature of $\qty{300}{K}$.
The increasing energy barrier at high stress magnitudes in the backward direction is qualitatively consistent with the reduced dislocation velocity shown in Fig.~\ref{fig:mobility}(a); the differences in stress magnitudes 
%\textcolor{red}{of the peak of the energy barriers}
are caused by the different simulation cell sizes used in the two (MD and NEB) calculations.
Thus, the observed mobility asymmetry in the jogged dislocation is connected to the non-monotonic dependence of the energy barrier on the applied stress in the backward direction.
What is the cause of this counterintuitive behavior?
%
% (moved to later)
%The answer lies in the coupling between the eigenstrain $\boldsymbol{\gamma}^*$ and vector $\boldsymbol{a}$.
%
%We note that in usual stress-driven thermally activated processes, the applied stress is solely coupled with the eigenstrain and no vector $\boldsymbol{a}$ is involved.

We start the explanation by first looking at the ``normal'' behavior of the forward transition (see Fig.~\ref{fig:mechanism}), where \emph{State L} goes to \emph{State R} assisted by a positive stress $\tau_{x'y}$.
For this to happen, $\gamma^{*}_{x'y}$ needs to turn positive and $a_{x'}$ needs to turn negative.  
The positive applied stress $\tau_{x'y}$ drives $\gamma^{*}_{x'y}$ in the positive direction, and, through the coupling between $\boldsymbol{\gamma}^{*}$ and $\boldsymbol{a}$, also exerts a force on $\boldsymbol{a}$ in the $-x'$-direction.
We shall call this force on the core atom the elastic force, $F_{\rm el}$, as depicted in the right inset in Fig.~\ref{fig:energy_barrier}.
The core atom also experiences a direct force, called the core force $F_{\rm c}$, through the short-range interactions with the stacking layers immediately above and below it.
As discussed earlier, while in \emph{State L}, the core atom interacts more strongly with the atoms in \emph{Layer A} below it.
This is indicated by a green vertical bar below the core atom in Fig.~\ref{fig:mechanism}(c) and Fig.~\ref{fig:energy_barrier}.
The application of a positive $\tau_{x'y}$ moves \emph{Layer A} to the left, producing a negative $F_{\rm c}$ on the core atom, as shown in the right inset of Fig.~\ref{fig:energy_barrier}.
Therefore, in the forward loading direction $\tau_{x'y} > 0$, both $F_{\rm el}$ and $F_{\rm c}$ point in the same direction, pushing the core atom to the $-x'$-direction needed to transition to \emph{State R}.

We now examine the ``abnormal'' behavior of the backward transition (see Fig.~\ref{fig:mechanism}), where \emph{State R} goes to \emph{State L} assisted by a negative stress $\tau_{x'y}$.
For this to happen, $\gamma^{*}_{x'y}$ needs to turn negative and $a_{x'}$ needs to turn positive.  
The negative applied stress $\tau_{x'y}$ drives $\gamma^{*}_{x'y}$ in the negative direction, and, through the coupling between $\boldsymbol{\gamma}^{*}$ and $\boldsymbol{a}$, also exerts a positive force $F_{\rm el}$
%\textcolor{red}{WHy positive?} 
on the core atom, as depicted in the left inset in Fig.~\ref{fig:energy_barrier}.
However, while in \emph{State R}, the core atom interacts more strongly with the atoms in \emph{Layer C} above it.
This is indicated by a blue vertical bar above the core atom in Fig.~\ref{fig:mechanism}(c) and the left inset of Fig.~\ref{fig:energy_barrier}.
The application of a negative $\tau_{x'y}$ moves \emph{Layer C} to the left, producing a negative $F_{\rm c}$ on the core atom, as shown in the left inset of Fig.~\ref{fig:energy_barrier}.
Therefore, in the backward loading direction $\tau_{x'y} < 0$, $F_{\rm c}$ points in the direction opposite to $F_{\rm el}$, and counteracts the push on the core atom towards the $+x'$-direction needed to transition to \emph{State L}.
The effect of the core force $F_{\rm c}$ becomes dominant over that of $F_{\rm el}$ at higher stress magnitudes, causing the energy barrier to increase and the dislocation velocity to drop.
\wei{To help build intuition, we sketch a mechanical device that illustrates the working of this ratchet mechanism in Supplementary Fig.~S10.}

We thus conclude that the cause of the forward-backward mobility asymmetry of jogged dislocations is intimately connected to the coupling between the eigenstrain $\boldsymbol{\gamma}^*$ (associated with dislocation motion) and vector $\boldsymbol{a}$ (associated with core atom displacement).
We note that in ``normal'' stress-driven thermally activated processes (including dislocation motion), the applied stress is solely coupled with the eigenstrain and no vector $\boldsymbol{a}$ is involved.

% \subsection{Stress dependence of mobility asymmetry}
% \textcolor{red}{How to edit this subsection? DO we need to move them to supplementary materials?}
\subsection{Generality and implications of dislocation mobility asymmetry}

To assess the generality of this mobility asymmetry, we performed additional MD simulations.
First, we repeated the MD simulations of jogged $\qty{30}{\degree}$ dislocations using another interatomic potential model for FCC nickel~\cite{Rao1999pma}, and observed qualitatively the same behavior as discussed above~{(see Supplementary Fig.~S4)}. 
%\hl{(Wei: I have edited SI tex file.  Please re-compile and re-upload SI pdf.  Potential ``used'' or ``developed'' by Rao.)}
%
Second, we examined dislocations with other character angles.
Jogged 60$^\circ$ dislocations also exhibited forward-backward mobility asymmetry~{(see Supplementary Fig.~S5)}. 
However, perfect edge, i.e. 90$^\circ$ dislocations do not exhibit mobility asymmetry, even when they contain jogs~{(see Supplementary Fig.~S6)}.
These results suggest that mobility asymmetry is a general behavior of jogged dislocations in FCC crystals except for dislocations with special orientations.

Experimental confirmation of the asymmetric dislocation mobility predicted here could be achieved by \textit{in situ} observations, such as transmission electron microscopy (TEM) or dark field X-ray microscopy (DFXM)~\cite{dresselhaus-marais_situ_2021}.
%\hl{[Refs]}.
%
The jogged dislocations can be produced by prior plastic deformations along specific loading orientations, which induce dislocation intersections between selected slip planes.
The cyclic stress can be applied by a loading gauge similar to a high-cycle fatigue test, or by a transducer for generating ultrasound, provided that the applied stress amplitude is in the intermediate range ($\qtyrange{10}{100}{MPa}$).
The predicted asymmetric mobility would lead to net uni-directional dislocation motion under the zero-mean cyclic loading.

On the other hand, the accumulation of plastic strain under cyclic loading, known as ratcheting (cyclic creep), is a critical failure mechanism in metallic structural materials. 
Ratcheting usually happens when the cyclic loading is asymmetric and has a non-zero mean stress, which induces continuous plastic strain in one direction due to dislocation motion~\cite{pan_superior_2025}.
Our findings on the asymmetrical mobility of jogged dislocations even under symmetric loading provide new insights into ratcheting, as well as into fatigue damage evolution in general.
The discovery of jog-induced asymmetric dislocation mobility challenges classical descriptions of plastic deformation in metals and alloys, and opens up new possibilities for micro-structural defect engineering in crystalline materials.

% The accumulation of plastic strain under cyclic loading, known as ratcheting (cyclic creep), is a critical failure mechanism in metallic structural materials.
%
% Ratcheting happens when the cyclic loading is asymmetric and has a non-zero mean stress, which induces continuous plastic strain in one direction, fundamentally governed by the collective motion of dislocations [Pan et al., Science, 2025].
%
% This finding is fundamentally important for understanding the plastic deformation of crystalline materials (including the majority of metals and alloys) and provide a new atomic-scale mechanism that directly influences fatigue behavior under cyclic stress.
%
% This ratchet mechanism provides a new framework for micro-structural engineering under cyclic loading conditions for designing materials with exceptional fatigue-resistance.

% ******************** 2025/11/26 ********************

% \hl{(Possible connecting with cyclic creep, ratcheting? Yifan please add)}

\section{Methods}
\label{sec:methodology}
\subsection{Simulation Tool and Interatomic Potential}
Molecular dynamics (MD) simulations of dislocation motion in single crystal nickel are conducted using the LAMMPS simulation package~\cite{thompson2022cpc}.
The interatomic interactions of Ni atoms are described by the embedded-atom method (EAM) interatomic potential developed by Angelo et al.~\cite{angelo1995msmse}. This potential was also used successfully by many previous works to simulate dislocation jogs~\cite{rodney2000prb, wang2025scripta, jian2026acta}. 

\subsection{Molecular Dynamics (MD) Simulation}
The schematic in Fig.~\ref{fig:mobility}(a) illustrates the simulation geometry setup of a mixed $\qty{30}{\degree}$ dislocation used to calculate the dislocation mobility.
The simulation cell size is about $\qtyproduct{240x150x360}{\angstrom}$ in the $[1\bar{1}0]\times[111]\times[\bar{1}\bar{1}2]$ directions.
Periodic boundary conditions were applied along the $x$- and $z$- directions while the $y$- direction had free surfaces.
A straight dislocation was first introduced to the crystal, and a unit jog pair was subsequently introduced by removing a row of atoms immediately above the line and shearing the atoms near them. %\hl{(see Supplementary Fig.~XX)}. 
The resulting atomic configuration was first relaxed to a local energy minimum and then equilibrated to a temperature of $T = \qty{300}{K}$ using MD simulations.
To drive dislocation motion, we applied a shear stress $\tau$ via forces on the surface atoms along the Burgers vector direction.
The arrows of $\tau$ (in Fig.~\ref{fig:mobility}) correspond to the forward direction, which causes the jogged dislocation to move in the $+x$-direction.
The instantaneous dislocation position during the MD simulation was determined using the dislocation analysis (DXA) algorithm in the OVITO software package~\cite{stukowski2012msmse}.
The dislocation velocity was extracted from the position-time curve from long MD simulations after steady-state motion was observed (see {Supplementary Fig.~S3}).

\subsection{Nudged Elastic Band (NEB) Calculation}
To improve computational efficiency, energy barrier calculations were performed using a reduced simulation cell of $\qtyproduct{49x47x52}{\angstrom}$ oriented along the $[1\bar{1}0]$, $[111]$, and $[\bar{1}\bar{1}2]$ directions.
The reduced geometry preserves the mobility asymmetry observed in larger-scale MD simulations ({see Supplementary Fig.~S2}).
The minimum-energy paths (MEP) for jog migration were determined using the nudged elastic band (NEB) method~\cite{henkelman2000jcp} as implemented in the LAMMPS software~\cite{thompson2022cpc}.
The initial and final configurations (denoted as \emph{State L} and \emph{State R}) were extracted from MD trajectories and subsequently relaxed via energy minimization.
These states represent two adjacent energy minima corresponding to a jog displacement of exactly one Burgers vector.
Stress-driven NEB calculations were conducted by applying a stress $\tau$ to the atoms on the free surfaces in the $y$-direction.

% \textcolor{red}{WJ: Yifan, can you write something about NEB details? Add force to the free surfaces while doing NEB}

\newpage

\textbf{Acknowledgements}

This work was supported by the National Science Foundation under Award Number DMREF 2118522 (W.J. and W.C.).
Y.W. was supported by the Stanford Energy Postdoctoral Fellowship and the Precourt Institute for Energy.

\bibliographystyle{elsarticle-num-names}
\bibliography{ref.bib}

% update supplementary PDF


\section*{Supplementary material}

# Supplementary material for “Forward-Backward Asymmetry in the Mobility of Jogged Dislocations in Face-Centered Cubic Nickel”

Wu-Rong Jian[a,1], Yifan Wang[a,1], Wei Cai[a,*]

[a]*Department of Mechanical Engineering, Stanford University, Stanford CA, 94305, USA*

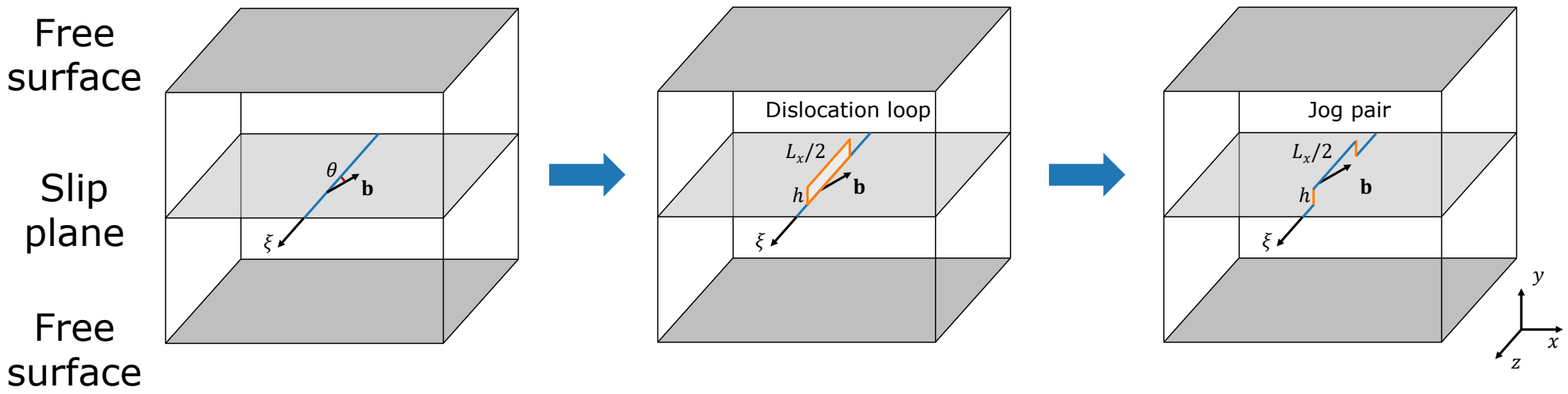


Figure S1: Schematic of the process for creating a unit jog pair on a straight dislocation in single-crystal nickel with two free surfaces along the $y$-direction. A straight dislocation with Burgers vector **b** and character angle $\theta$ is first introduced into the crystal. Subsequently, a dislocation loop, also with Burgers vector **b**, is generated such that it partially overlaps with the existing straight dislocation. The overlapping segment undergoes mutual annihilation, resulting in the formation of a jog pair along the dislocation line.

*Corresponding author
*Email address:* `caiwei@stanford.edu` (Wei Cai)
[1]Both authors contributed equally to the paper

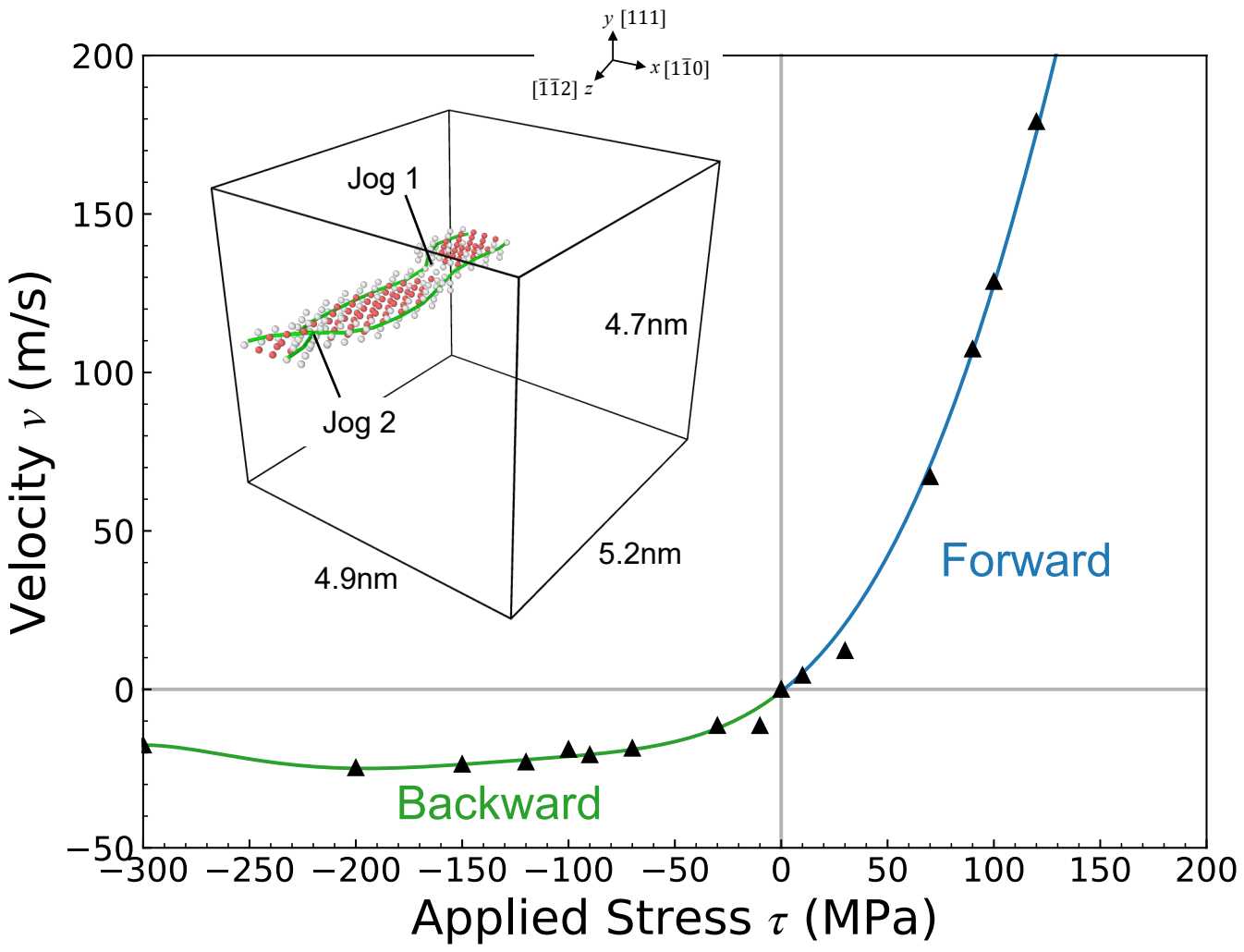


Figure S2: Asymmetry of the jogged dislocation mobility observed in the smaller simulation cell used for the nudged-elastic band (NEB) calculation of energy barriers. The inset is the relaxed configuration of the small simulation cell with the size of 49 Å×47 Å×52 Å oriented along the [1$\bar{1}$0], [111], and [$\bar{1}\bar{1}$2] directions.

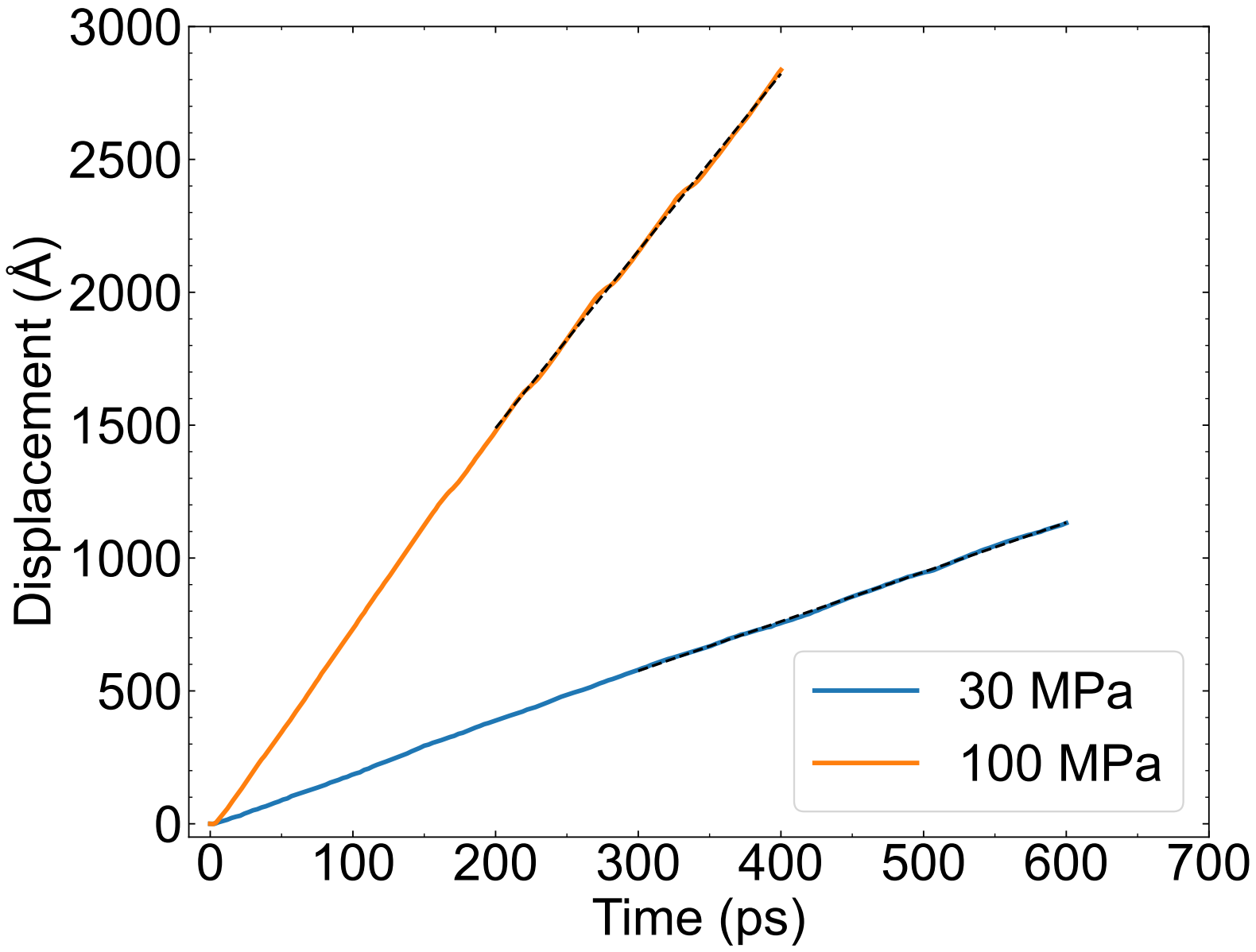


Figure S3: Dislocation displacement as a function of time for the jogged dislocations with character angle $\theta = 30°$ at 300 K, under shear stresses of 30 MPa and (b) 100 MPa, respectively. Molecular dynamics simulations are performed using the Angelo et al. (1995) EAM potential. The dislocation velocity is determined from the slope of the second half of simulation period, after the dislocation motion has reached a steady-state.

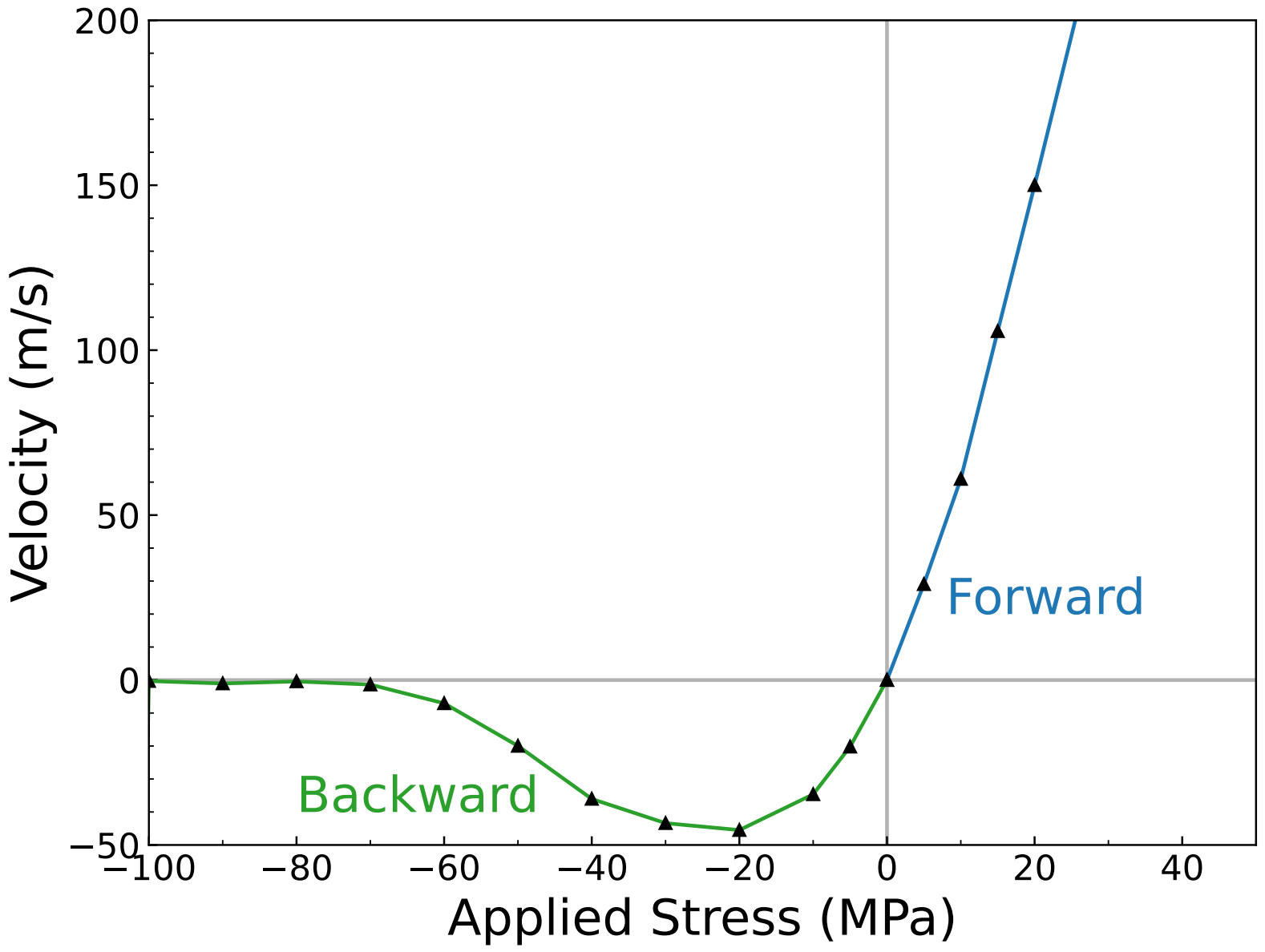


Figure S4: Velocity of jogged dislocation in FCC nickel with character angle $\theta$ = 30° as a function of applied stress $\tau$. The results are simulated using the potential developed by Rao et al. (1999).

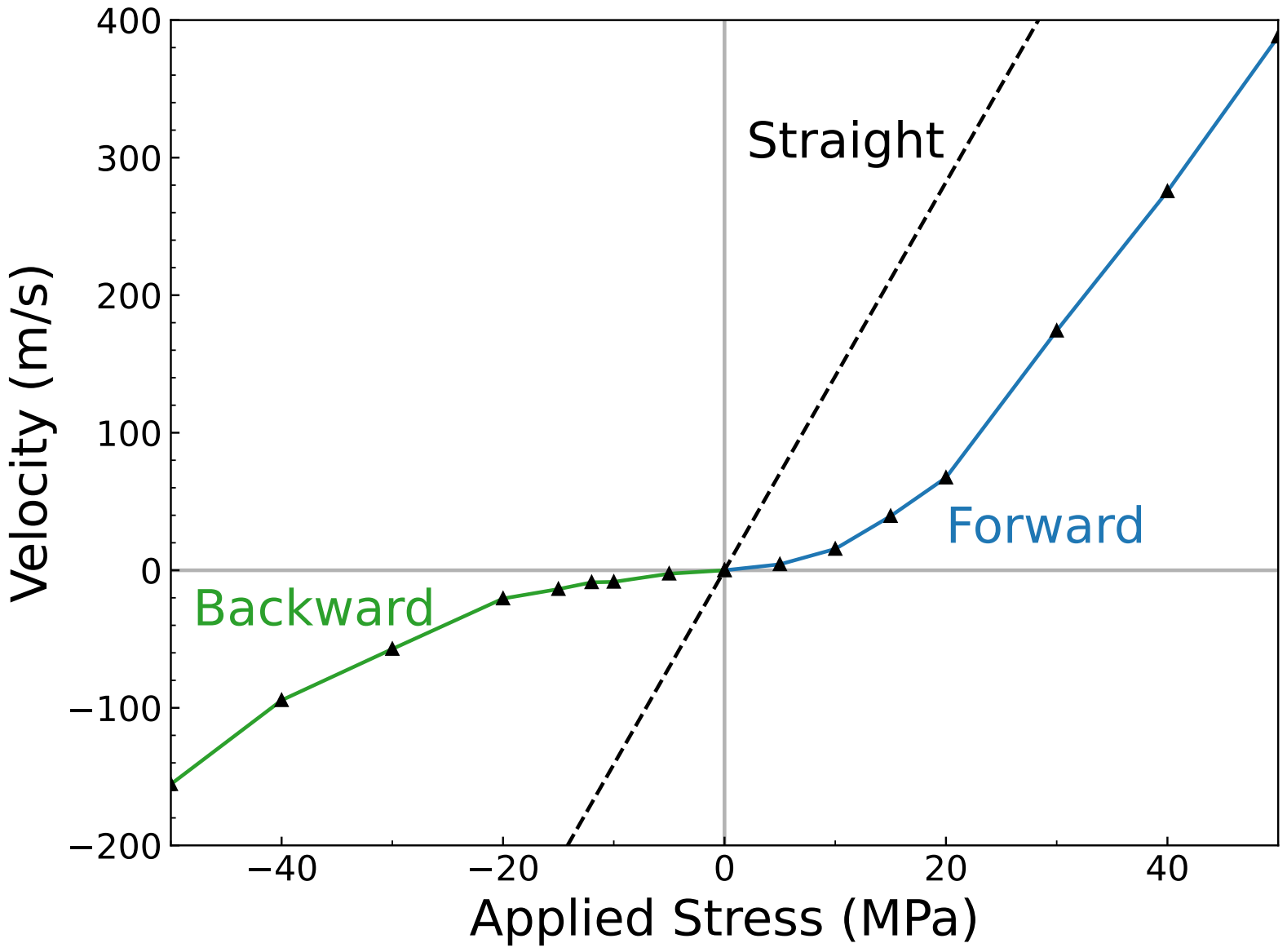


Figure S5: Velocity of jogged (solid line) and straight (dashed line) dislocations with character angle $\theta$ = 60° as a function of applied stress $\tau$. The results are simulated using the Angelo et al. (1995) EAM potential.

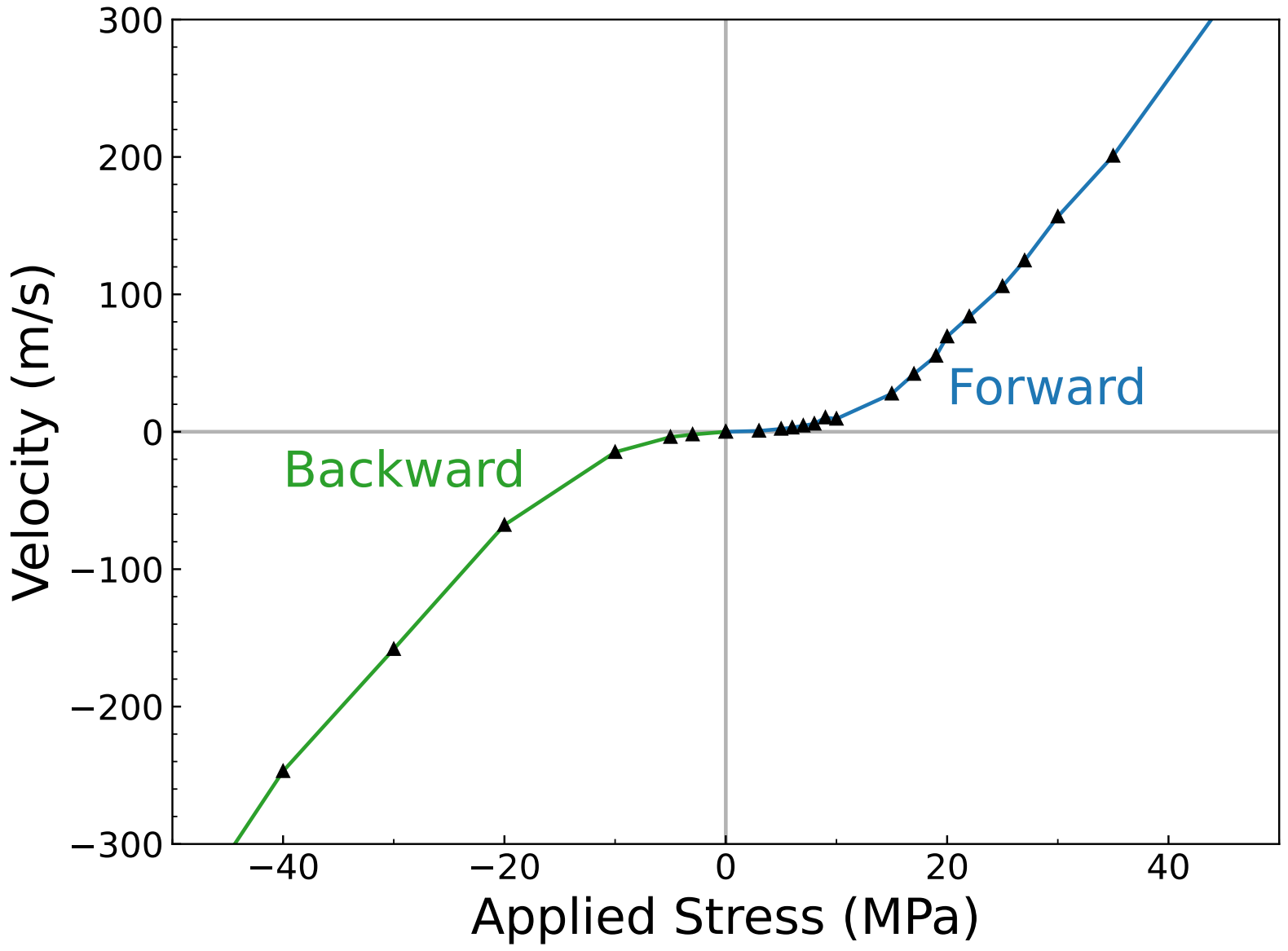


Figure S6: Velocity of jogged dislocation with character angle $\theta = 90°$ (edge dislocation) as a function of applied stress $\tau$. The results are simulated using the Angelo et al. (1995) EAM potential.

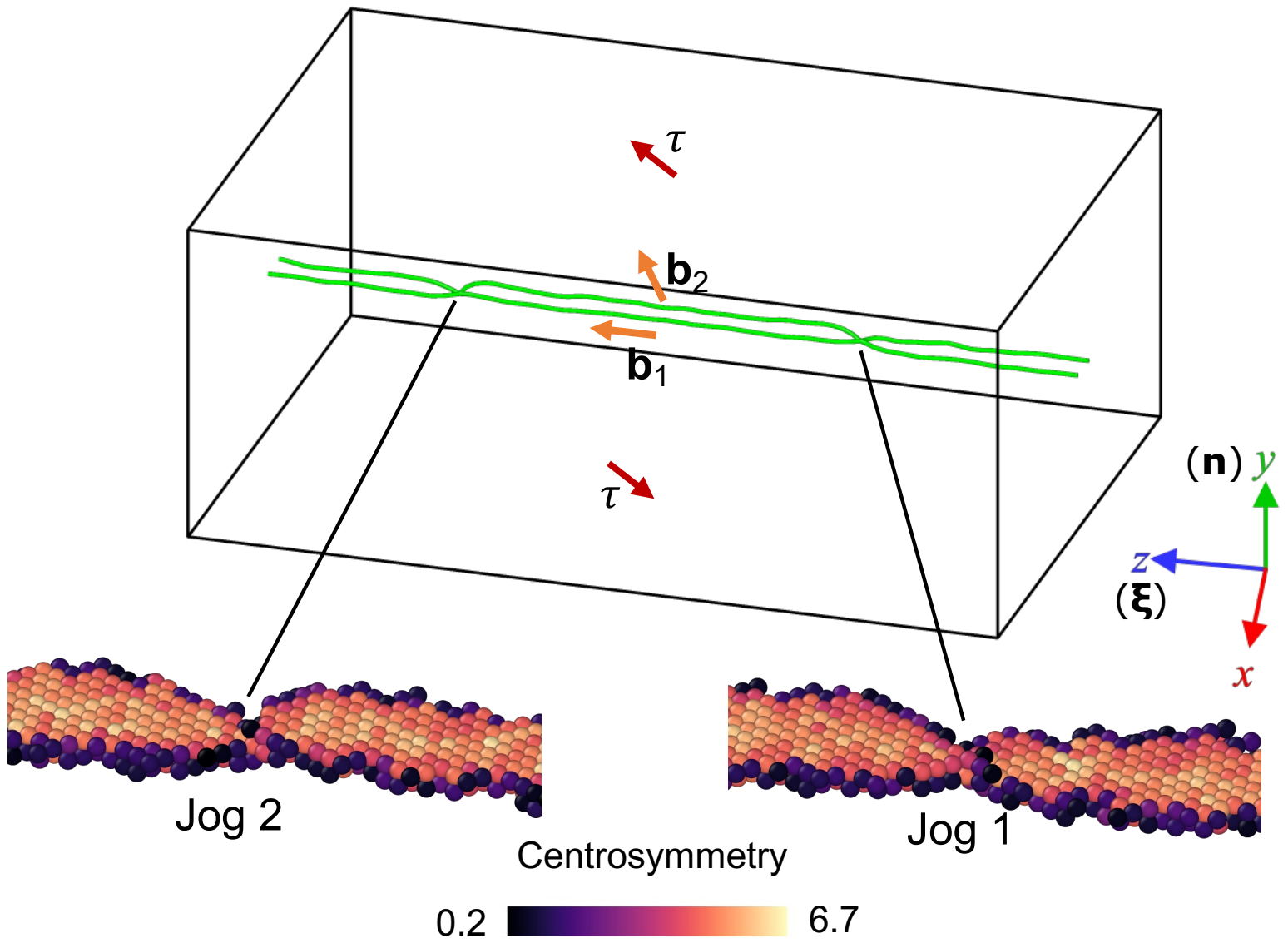


Figure S7: (a) Relaxed dislocation configuration with a unit jog pair and character angle $\theta = 30°$. The Burgers vector $\mathbf{b}$ and the line vector $\boldsymbol{\xi}$ follows the right-hand, start-finish (RHSF) convention used in Cai and Nix (2016). The dislocation segments on the slip plane dissociate into Shockley partials (orange arrows) $\mathbf{b}_1$ and $\mathbf{b}_2$. The atoms are colored with the centrosymmetry parameter.

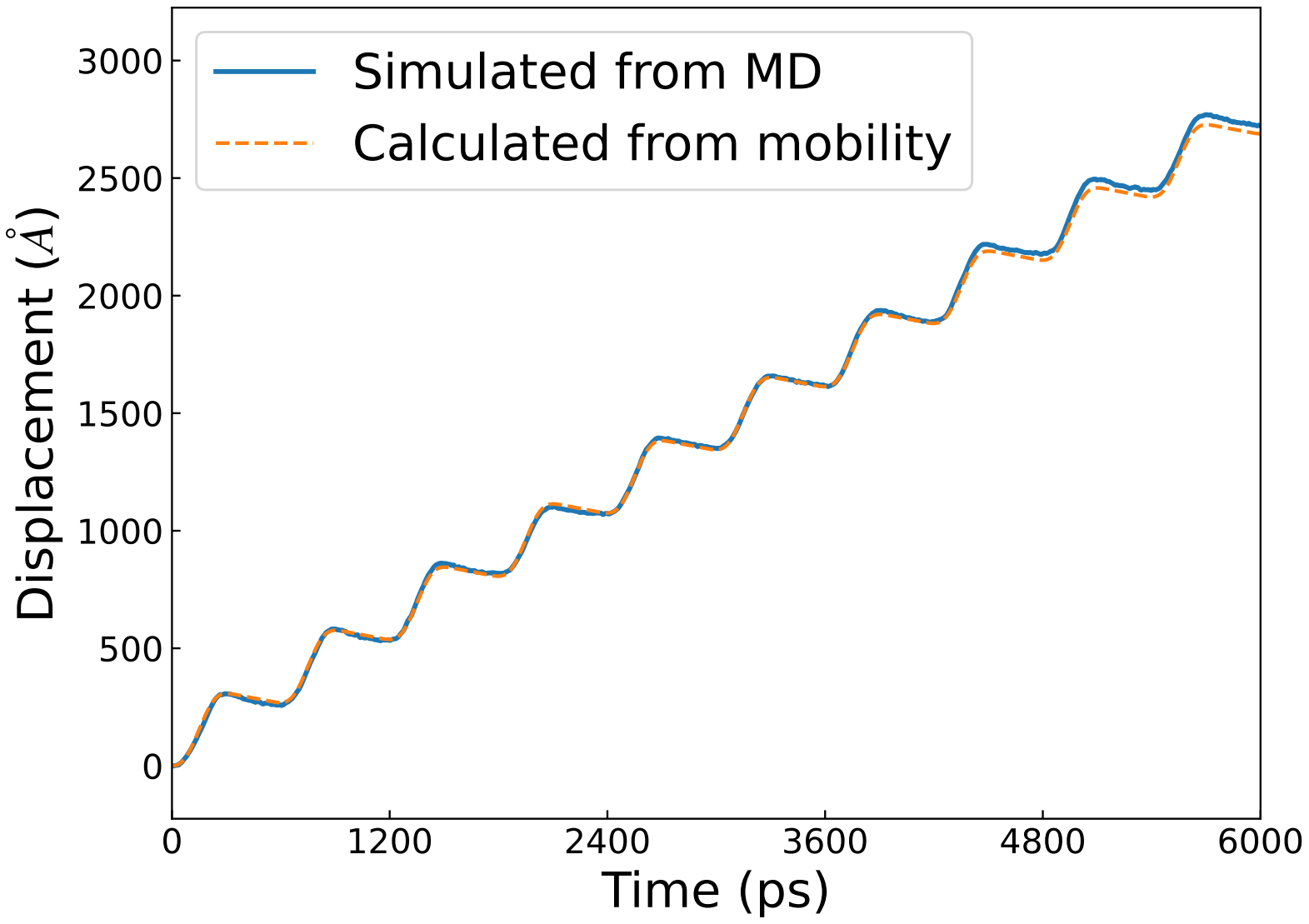


Figure S8: Additional displacement-time curve of the dislocation motion under sinusoidal cyclic loading at a stress amplitude of 30 MPa. The blue solid line indicates the MD simulation result of the displacement. The orange dashed line shows the calculated displacement-time curve, as the integration of the intantaneous velocity of the velocity-stress mobility curve.



\end{document}